\documentclass[reprint,prl,aps,superscriptaddress,amsmath,amssymb,longbibliography]{revtex4-1}
\usepackage{graphicx}% Include figure files
\usepackage{dcolumn,tabularx}% Align table columns on decimal point
\usepackage{bm}% bold math
\usepackage{braket}

\usepackage{mathdots}
\usepackage[T1]{fontenc}
\usepackage[unicode=true]{hyperref}
\usepackage{booktabs}
\usepackage{mathtools}
\usepackage{color}

\begin{document}
	%\linenumbers
	
	\title{Observation of Non-Markovian Evolution of Tripartite Quantum Steering}
	
	\author{Yan Wang}
	\affiliation{School of Physics, Hangzhou Normal University, Hangzhou 310036, China}	
	
	\author{Shao-qi Lin}
	\affiliation{School of Physics, Hangzhou Normal University, Hangzhou 310036, China}
	
	\author{Rui-qi Shen}
	\affiliation{School of Physics, Hangzhou Normal University, Hangzhou 310036, China}
	
	\author{Fang-liang Chen}
	\affiliation{School of Physics, Hangzhou Normal University, Hangzhou 310036, China}
	
	\author{Guo-qiang Zhang}
	\affiliation{School of Physics, Hangzhou Normal University, Hangzhou 310036, China}
	
	\author{Li-jiong Shen}
	\affiliation{School of Physics, Hangzhou Normal University, Hangzhou 310036, China}
	
	\author{Yong-nan Sun}
	\email{synan@hznu.edu.cn}
	\affiliation{School of Physics, Hangzhou Normal University, Hangzhou 310036, China}
	
	\author{Qi-ping Su}
	\email{sqp@hznu.edu.cn}
	\affiliation{School of Physics, Hangzhou Normal University, Hangzhou 310036, China}
	
	\author{Chui-ping Yang}
	\email{yangcp@hznu.edu.cn}
	\affiliation{School of Physics, Hangzhou Normal University, Hangzhou 310036, China}

	\begin{abstract}		 
		%In open quantum systems, the investigations of multipartite Einstein-Podolsky-Rosen (EPR) steering evolution holds profound significance for advancing practical quantum technologies, which bridges the gap between idealized theoretical models and real-world quantum network implementations. Particularly, the non-Markovian environment, where memory effect plays an important role, would govern the information backflow that drives the multi-stage revival of steering correlations. 
		%Here, we observe the non-Markovian evolution of tripartite steering including the death and revival processes, and introduce the new method to quantify the degree of non-Markovianity through the whole evolution process. Our findings provide foundational insights for harnessing rich hierarchy correlations of multipartite steering resources in distributed quantum computing and quantum networks under realistic noise, and offer the potential for robust quantum technology applications.
		
		The memory effects in open quantum systems can induce information backflow and revive quantum correlations, thereby providing a powerful way to protect and recover useful quantum resources in realistic noisy environments. However, such dynamics remains experimentally unexplored in multipartite quantum steering. Here we observe different non-Markovian evolution of tripartite quantum steering using Greenberger-Horne-Zeilinger-type mixed states, covering both death and revival processes. In particular, we experimentally demonstrate the more intricate asymmetric steering structure of tripartite quantum steering through different bipartitions, which do not arise in bipartite systems. Our results provide foundational insights into the hierarchical and directional structures in multipartite quantum steering, and highlight its potential as a useful resource for asymmetric quantum information processing.	
	\end{abstract}
	
	%Additionally, the complete death and revival evolution of quantum correlations also offer an experimentally accessible method of quantifying the degree of non-Markovianity.
	% we also quantify the degree of non-Markovianity through the complete death and revival evolution of quantum correlations.
	
	\maketitle
	
	%理论部分加一个非马尔可夫程度的数学表达式+去掉提出新方法的claim
	
	\emph{Introduction.}---Open quantum systems \cite{RevModPhys.88.021002}, where unavoidable interactions with the environment cause decoherence and dissipation, are fundamental to understanding quantum information science in realistic scenarios \cite{harrington2022engineered}. Unlike idealized closed systems, open systems generally experience irreversible information loss, typically described by Markovian processes \cite{breuer2002theory,4frd-ck2z}. By contrast, non-Markovian processes, characterized by memory effects that allow information flow back to the system, can induce counterintuitive phenomena such as the sudden death and revival of quantum correlations \cite{PhysRevLett.101.150402,PhysRevX.10.041049,RevModPhys.89.015001,shen2026non}. These memory-driven dynamics are critical for quantum technologies in noisy environments \cite{PhysRevLett.104.100502,PhysRevA.82.022324,liu2011experimental,liu2013photonic,xu2013experimental,liu2018experimental,PhysRevA.102.062208,PhysRevA.108.012213}, enabling preservation and robust conservation of quantum resources \cite{mmy1-jb4n} and suppressed dissipation  \cite{wfyl-wtz3}. Therefore, studying the non-Markovian process in multipartite quantum systems is necessary, which would reveal more complex hierarchical structures among subsystems and provide foundational applications in scalable quantum computation \cite{main2025distributed,wk87-5vnv} and quantum networks  \cite{RevModPhys.92.015001,PhysRevLett.125.020404}.
	
	Quantum steering, first conceptualized by Schrödinger \cite{1935,1936} in response to the Einstein-Podolsky-Rosen paradox \cite{epr}, describes the ability of one observer to remotely influence another system's state through local measurements. In 2007, it was rigorously defined with experimental criteria \cite{PhysRevLett.98.140402}, and reformulated as a  distinct category of nonlocality between entanglement \cite{RevModPhys.81.865} and Bell non-locality \cite{RevModPhys.86.419}. %Taking bipartite scenarios (Alice and Bob) as example, quantum steering from Alice to Bob is demonstrated if Bob's assemblage of conditional states cannot be  explained by a local-hidden-state model \cite{RevModPhys.92.015001}. 
	Crucially, a distinct feature of steering is its  asymmetry property  \cite{one-wayex-2016,gehring2015implementation}, %which allows one-way steering, namely, Alice can steer Bob while Bob cannot steer Alice  
	which underpins important one-sided device-independent (1SDI) tasks such as quantum key distribution \cite{bouwmeester1997experimental,PhysRevA.85.010301}, randomness certification \cite{PhysRevLett.132.080201}, and subchannel discrimination \cite{gehring2015implementation,sun2018sub}.  When extended from bipartite to multipartite scenarios, the more complex hierarchical structures would exhibit genuinely many-body nonclassical correlations, for example, shareability \cite{PhysRevLett.128.120402,cai2025review} and collective steering \cite{PhysRevLett.111.250403,PhysRevA.105.012202,PhysRevA.111.032415}. Moreover, multipartite quantum steering has been experimentally demonstrated in different systems \cite{PhysRevA.84.032115,PhysRevLett.111.250403,armstrong2015multipartite,PhysRevLett.118.230501,kunkel2018spatially,PhysRevA.95.010101,PhysRevA.99.012302}, significantly broadening its relevance to large-scale quantum networks \cite{PhysRevLett.125.260506}. Taking tripartite systems as an example, the degree of steerablity  manifests in two distinct scenarios, i.e., 1SDI and two-sided device-independent (2SDI) \cite{cavalcanti2015detection}. In 1SDI scenario, only one party's measurement apparatus is untrusted while the other two parties' are trusted. In 2SDI scenario, two parties' measurement devices are unstrusted and the remaining one is trusted. 
	Although plenty of work in multipartite quantum steering has been investigated \cite{RevModPhys.92.015001,cai2025review}, the dynamical behaviors in non-Markovian environments, where memory effects play a central role, still remain largely unexplored.
	
	In this work, we study distinct tripartite quantum steering evolution in the non-Markovian environment using Greenberger-Horne-Zeilinger (GHZ)-type mixed states, which incorporates the asymmetric noise to reveal the directional property in tripartite scenarios. During system-environment interactions, we observe the complete dynamical evolution of the system state in 1SDI and 2SDI scenarios through different configurations of subsystems. And we experimentally demonstrate the intricate asymmetric structure in tripartite quantum steering. The system state, initially steerable in 1SDI and 2SDI scenarios, progressively evolves into steerable only in 2SDI scenario and then unsteerable in both scenarios. Subsequently, due to the memory effect, it revives back into steerable only in 2SDI scenario and eventually steerable in two scenarios. %We also provide the new method of characterizing the degree of non-Markovianity through the death and revival of quantum correlation. 
	Our results demonstrate the asymmetric nature of multipartite quantum steering in open quantum systems, and take a step forward in studying potential applications in noisy quantum networks.

	\begin{figure}[t]
		\includegraphics[width=0.48\textwidth]{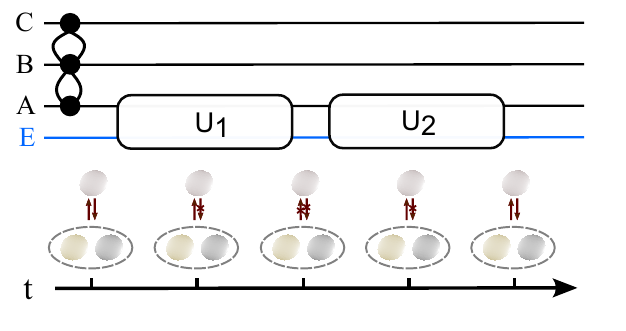}
		\caption{\textbf{Theoretical illustration of ideal tripartite non-Markovian evolution.}  The subsystem (A) of tripartite state (composed of A, B and C) undergoes the time-dependent interactions (varying with t) with its surrounding environments (E) denoted with operations $\mathrm{U}_1$ and $\mathrm{U}_2$. The whole tripartite quantum steering dynamical evolution, including the sudden death and revival processes, covers steerable in one-sided and two-sided device-independent (1SDI and 2SDI) scenarios, only steerable in 2SDI scenario, and unsteerable in both scenarios. 
		}\label{th}
	\end{figure}

	\emph{Theoretical analysis of non-Markovian evolution in GHZ-type mixed states.}---Within the open quantum system framework, the system interacts with its surrounding environment, and the initial global state at $\mathrm{t}=0$ is assumed to be factorized as $\rho_g(\eta, \theta) = \rho_{\mathrm{ABC}}^s(\eta, \theta) \otimes \rho_e$, where $\rho_{\mathrm{ABC}}^s(\eta, \theta)$ and $\rho_e$ denote the system and environment states \cite{PhysRevLett.130.200202}. 
	In our photonic implementation, the system state is encoded in the photon's path and polarization degrees of freedom, while the environment is introduced through the photon's frequency degree of freedom. Therefore, the system state $\rho_{\mathrm{ABC}}^s(\eta, \theta)$ is taken to be a tripartite GHZ-type mixed state, constructed from the coherent superposition $\cos\theta\ket{000}+\sin\theta\ket{111}$ together with an additional noisy admixture, in the form 
	\begin{equation}
		\label{eq1}
		\rho_{\mathrm{ABC}}^s(\eta,\theta)=\eta\ket{\Phi(\theta)}\bra{\Phi(\theta)}+(1-\eta)\openone_A\otimes\rho_{BC}^\theta/2,
	\end{equation}
	where $\ket{\Phi(\theta)}=\cos\theta\ket{000}+\sin\theta\ket{111}$, $\rho_{BC}^{\theta}={\rm Tr}_{A}[\ket{\Phi(\theta)}\bra{\Phi(\theta)}]$ and $\openone_A$ is the identity operator on subsystem $A$. As shown in Fig. \ref{th}, the time-dependent operations $\mathrm{U}_1$ and $\mathrm{U}_2$ are successively applied on subsystem A. The first process $\mathrm{U}_1$ induces the loss of system information to environment and degrades the degree of quantum correlation, thereby obtaining the reduced density matrix of system state 
	\begin{equation}
		\begin{aligned}
			\label{eq2}
			\rho&_{\mathrm{ABC}}^s(\eta,\theta,t)=\\%0.5*\\
			%&\left(
			%\begin{array}{ccccccccc}
			%	(1+\eta)\cos^2\theta & 0  & 0  & 0  & 0  & 0  & 0  & \kappa_a(t)\eta\cos\theta\sin\theta \\
			%	0 & 0 & 0 & 0 & 0 & 0 & 0 & 0\\
			%   0 & 0 & 0 & 0 & 0 & 0 & 0 & 0\\
			%	0 & 0 & 0 & (1-\eta)\cos^2\theta & 0 & 0 & 0 & 0\\
			%	0 & 0 & 0 & 0 & (1-\eta)\sin^2\theta & 0 & 0 & 0\\
			%	0 & 0 & 0 & 0 & 0 & 0 & 0 & 0\\
			%	0 & 0 & 0 & 0 & 0 & 0 & 0 & 0\\
			%	\kappa_a^*(t)\eta\cos\theta\sin\theta & 0  & 0  & 0  & 0  & 0  & 0  & (1+\eta)\sin^2\theta \\
			%\end{array}\right).	\\
			&[(1+\eta)\cos^2\theta\, |000\rangle\langle000| + (1+\eta)\sin^2\theta|111\rangle\langle111| \\
			&+(1-\eta)\cos^2\theta\, |011\rangle\langle011| + (1-\eta)\sin^2\theta\, |100\rangle\langle100| \\
			&+\kappa_1(t)\eta\cos\theta\sin\theta(|000\rangle\langle111| + |111\rangle\langle000|)]/2.
		\end{aligned}
	\end{equation}
	To simplify the model, we assume that Bob and Charlie share the same environmental frequency, i. e., $\omega_b=\omega_c=\omega_2$, while Alice is associated with $\omega_a=\omega_1$. Therefore, the above decoherence function is $\kappa_1(t=t_1)=\int d\omega_1 d\omega_2 F(\omega_1,\omega_2)\exp(-i\omega_1 t_1)$, in which $t_1$ is the dephasing time on Alice's side and $F(\omega_1,\omega_2)$ is the joint frequency amplitude distribution. As the system-environment interaction accumulates on Alice's side, the degree of quantum steering in system state  $\rho_{\mathrm{ABC}}^s(\eta,\theta,t)$ would undergo the process of gradual decay to death. Correspondingly, the steering evolution can exhibit transitions from steerable in both 1SDI and 2SDI scenarios to steerable only in 2SDI scenario, and eventually unsteerable in both scenarios, as illustrated in Fig. \ref{th}. Subsequently, under the operation $\mathrm{U}_2$ with a compensatory effect, the above decoherence function becomes  $\kappa_2(t=t_{1}+t_{2})=\int d\omega_1 d\omega_2 F(\omega_1,\omega_2)\exp[-i(\omega_1 t_1+\omega_2 t_2)]$, leading to the revival of steerability. 
	
		\begin{figure}[b]
		\includegraphics[width=0.5\textwidth]{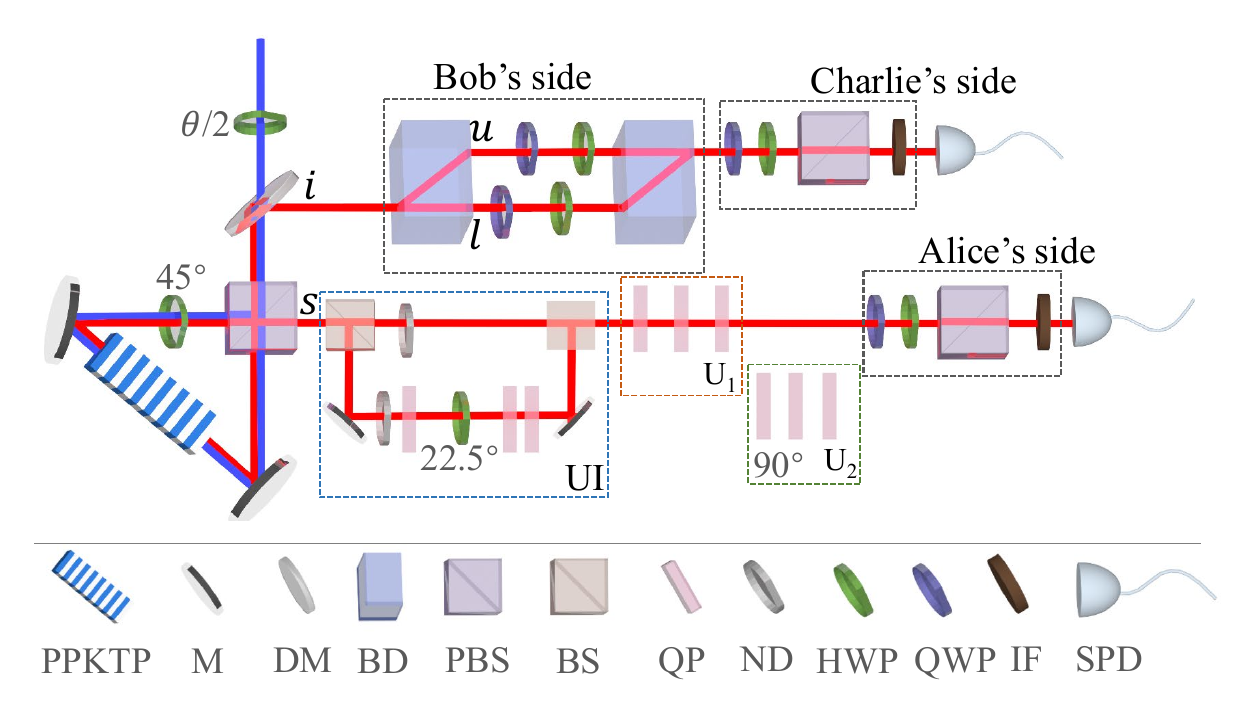}
		\caption{\textbf{Experimental setup.} Polarization-entangled photon pairs are generated via parametric down-conversion in a periodically poled potassium titanyl phosphate (PPKTP) crystal. One photon is sent to the interference which separated and combined by two beam displacers (BDs), while the other one passes through the unbalanced interferometer (UI) in the blue dashed box, preparing the initial system state $\rho_{\mathrm{ABC}}(\theta,\eta)$. On Alice's side, the sequence of quartz plates (QPs) at $0^\circ$ and $90^\circ$ are respectively inserted to introduce the time-dependent operations $\mathrm{U}_1$ and $\mathrm{U}_2$ corresponding to the orange and green boxes. We mark the QPs at $90^\circ$ in the green box, leaving the rest at $0^\circ$ in the blue and orange boxes unmarked. Three dotted boxes are the measurement apparatus including quarter-wave plate (QWP), half-wave plates (HWP), a BD or a polarized beam splitter (PBS). See the corresponding details in the main text. M: mirror, DM: dichroic mirror, BS: beam splitter, ND: netural density filters, IF: interference filter, SPD: single-photon detector. 
		}\label{expe}
	\end{figure}
	
	The recovery ability of steering is governed by the degree of non-Markovianity, which is denoted as $\mathcal{M}\in\{0,1\}$. Typically, the degree of non-Markovianity is quantified by
	\begin{equation}
		\begin{aligned}
			&\mathcal{N}
			=
			\max_{\rho_{1,2}(0)}
			\int_{\dot D(t)>0}\dot D(t)\,dt,\\
			&D(t)=\frac{1}{2}\mathrm{Tr}\left|\rho_1(t)-\rho_2(t)\right|,
		\end{aligned}
	\end{equation}
	which correlates with the maximal growth of distinguishability between two optimally chosen states $\rho_1$ and $\rho_2$ \cite{PhysRevLett.103.210401,PhysRevLett.108.210402,PhysRevLett.111.229901}. When $\mathcal{M} = 1$ (the ideal non-Markovian process), 
	the memory effect enables the system state evolves form unsteerable in 1SDI and 2SDI scenarios to steerable only in 2SDI scenario and steerable in both scenarios at the equal evolution time in $\mathrm{U}_1$ and $\mathrm{U}_2$ ($t_1=t_2$), as shown in Fig. \ref{th}. By contrast, $\mathcal{M}=0$ corresponds to a Markovian process without memory effects and hence no revival. Operationally, the degree of non-Markovianity can be determined from the experimentally observed death and revival dynamical evolution of quantum correlations, e.g., entanglement or quantum steering.

	%Likewise, the entanglement shared between Alice and Bob also undergoes sudden death and revival dynamic processes as interactions with the environment occur. 
	%Under the ideal non-Markovian system in Fig. \ref{th}, the revival dynamics are from two-way unsteerable to one-way steerable, until two-way steerable, when the evolution time in the operations of $\mathrm{\varepsilon_1}$ and $\mathrm{\varepsilon_2}$ is equal, i.e., $t_a=t_b$. 
	%The whole non-Markovian dynamics, including sudden death and revival processes, demonstrate the operation of $\mathrm{\varepsilon_2}$ can straightforwardly cancel the relative dephasing under the memory effect of the non-Markovian system.
	
	\emph{Experimental setup and results.}---To observe non-Markovian dynamical behaviors in multipartite systems, we characterize a tripartite state encoded in photons' path and polarization degrees of freedom. 
	As shown in Fig. \ref{expe}, a continuous-wave laser at 405 nm bidirectionally pumps a 10 mm-long Type-II periodically poled potassium titanyl phosphate (PPKTP) crystal in phase-stable Sagnac interference, generating the polarization-entangled states $\cos\theta\ket{H_sH_i}+\sin\theta\ket{V_sV_i}$, where the subscript $s$ ($i$) denote the signal (idler) photon, and $\theta$ is controlled by the half-wave plate (HWP) \cite{Fedrizzi:07}. 
	Here, $\ket{H}$ and $\ket{V}$ are horizontal and vertical polarizations. 
	The idler photon is then sent to a beam displacer (BD), which transmits the vertically polarized component along the lower path ($l$) and deflects the horizontally polarized component into the lower path ($u$) by 4 mm, thereby encoding polarization information into spatial modes. 
	The generated state is prepared in  $\ket{\Phi(\theta)}=\cos\theta\ket{H_sH_iu}+\sin\theta\ket{V_sV_il}=\cos\theta\ket{000}+\sin\theta\ket{111}$, where polarization $H$ ($V$) and path mode $u$ ($l$) are re-encoded $0$ ($1$) \cite{cavalcanti2015detection}. Meanwhile, the signal photon passes through an unbalanced interferemeter (UI), in which one path remains unchanged while the other one contains quartz plates (QPs) sandwiched by an HWP set at $22.5^\circ$. The inserted QPs, where the thickness of QPs placed after HWP is twice that of the QPs placed before the HWP, introduce sufficient decoherence between $H$ and $V$ components. By combining two paths into one, arbitrary GHZ-type mixed state $\rho_{\mathrm{ABC}}(\theta,\eta)$ can be prepared, with $\eta$ controlled by the neutral-density (NDs) filters. 
	
	\begin{figure}[t]
		\includegraphics[width=0.48\textwidth]{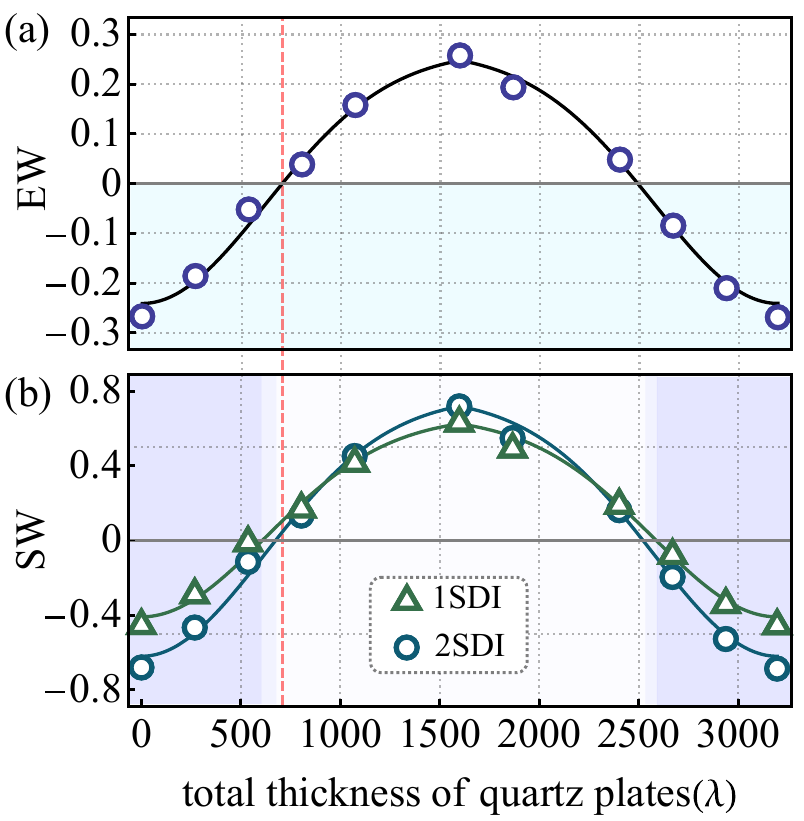}
		\caption{\textbf{The non-Markovian evolution processes of initial system state $\rho_{\mathrm{ABC}}(\theta=\pi/4,\eta=1)$.} (a) The evolution of entanglement witness (EW) versus the total thickness of quartz plates $\mathcal{L}(\lambda)$, with the entangled region in light blue region. (b) The evolution of steering  witness (SW) in one-sided and two-sided device-independent scenarios (1SDI and 2SDI) versus $\mathcal{L}(\lambda)$, with experimental results in hollow triangles and circles. The dark, light, and unshaded backgrounds are denoted the states steerable in 1SDI and 2SDI scenarios, steerable only in 2SDI scenarios, and unsteerable in both regions. The red dotted line denotes the entanglement boundary of $\mathcal{L}(\lambda)$. All error bars are calculated through the Monte Carlo method which are too small to be seen (the same applies to the following results). 
		}\label{fig-1}
	\end{figure}
	
	The time-dependent dephasing interaction is effectively engineered by tuning the optical-path difference between ordinary and extraordinary components, namely the effective thickness of QPs as $\mathcal{L} (\lambda)$. Accordingly, the time-dependent parameter $\kappa_1(t)$ is replaced by $\kappa_1(\mathcal{L}=\mathcal{L}_1)$ via $t_1=\mathcal{L}_1/c=\Delta n L_1/c$, where $\mathcal{L}_1$ is the effective optical path length on Alice’s side. Notably, $\Delta n$, $L_1$ and $c$ are the refractive difference of ordinary and extraordinary components, the physical thickness of QPs at $0^\circ$ on Alice's side and the speed of photons in vacuum. 
	Analogously, $\kappa_2(t)$ is written as $\kappa_2(\mathcal{L}=\mathcal{L}_1+\mathcal{L}_2)$, where $\mathcal{L}_2$ is the effective optical path of QPs at $90^\circ$ on Alice's side, corresponding to the green box in Fig. \ref{expe}. 
	As illustrated in Fig. \ref{expe}, the QPs at $0^\circ$ (in the orange box) are first inserted to implement system-environment operation $\mathrm{U}_1$, leading to the dephasing of non-locality correlation from steerable in 1SDI and 2SDI scenarios to steerable only in 2SDI scenario and eventually to unsteerable in both scenarios as shown in Fig. \ref{th}. 
	Due to the memory effect in non-Markovian evolution, the dephasing operation $\mathrm{U}_2$ introduced by QPs at $90^\circ$ (in the green box) revives the lost quantum correlation, where the system state evolves from unsteerable in both scenarios to steerable only in 2SDI and steerable in 1SDI and 2SDI scenarios as illustrated in Fig. \ref{th}.

	Then the signal photon is sent to Alice for polarization measurement via a quarter-wave plate (QWP), an HWP, and a polarized beam splitter (PBS). The idler photon is sent to Bob and Charlie, who perform measurements on the polarization and path degrees of freedom. On Bob's side, the combination of QWP, HWP, and BD enables both direct polarization measurement of the second qubit and the conversion of path information into polarization information. The final QWP, HWP, and PBS complete measurements of the path-encoded qubit. Afterward, the photons both pass through 3-nm-bandwidth interference filters (IFs) before being collected by single-photon detectors (SPDs). To fully characterize $\rho_{\mathrm{ABC}}(\theta,\eta)$, we perform standard three-qubit quantum state tomography via joint measurements in complementary bases to reconstructing the experimental density matrix $\rho_{ex}$ \cite{PhysRevA.64.052312}, with the average fidelity of $f=\mathrm{Tr}[\sqrt{\sqrt{\rho_{th}}\rho_{ex}\sqrt{\rho_{th}}}]=0.941(6)$ ($\rho_{th}$ is the theoretical density matrix). 
	
	%Conventional quantification of non-Markovianity relies on the death and revival processes of trace distance between two optimally chosen states	.
	Conventional quantification of non-Markovianity relies on the death and revival processes of trace distance between two optimally chosen states \cite{PhysRevLett.103.210401,PhysRevLett.108.210402,PhysRevLett.111.229901}. 
	Here, we introduce an operational approach to characterize the degree of non-Markovianity through detecting the whole death and revival evolution of quantum correlation. %Our method enables direct calculations of the degree of non-Markovian memory effects from experimentally obtained evolution. 	
	As illustrated in Fig. \ref{fig-1}, we first prepare the initial system state as $\rho_{\mathrm{ABC}}(\theta=\pi/4,\eta=1)$ and observe the non-Markovian evolution of tripartite entanglement (a) and quantum steering (b) by systematically varying the total effective thickness of QPs  on Alice's sides, namely $\mathcal{L}(\lambda)$, thereby controlling the strength of system-environment interactions.  
	During the death process governed by $\kappa_1$, the evolution could be written as the Gaussian function $g_1(x)=A*e^{-Bx^2}+C$, where $x$ is the induced effective thickness of QPs (at $0^\circ$), parameters $A$ and $B$ are related to the source parameters and the refractive index of QPs. Fitting the results yields the parameters as $A=0.507139$, $B=1.29992*10^{-6}$ and $C=0.266244$. Notably, the maximal thickness of QPs (at $0^\circ$) in the death process is marked as $\mathcal{L}_{max}$. The system-environment interaction governed by $\kappa_2$ is further applied through QPs at $90^\circ$ on Alice's side, then recovers the lost quantum resources. The corresponding revival ability is determined by the degree of non-Markovianity, denoted as $\mathcal{M}$. And the theoretical analysis of decoherence function is $g_2(x)=A*\exp(-B[\mathcal{L}_{max}^2+(x-\mathcal{L}_{max})^2-2\mathcal{M}*\mathcal{L}_{max}(x-\mathcal{L}_{max})])+C$. 
	From the reconstructed density matrix $\rho_{ex}$, the degree of entanglement is calculated through the tripartite entanglement witness (EW) as $\mathrm{Tr}[\rho_
	{ex}W]$, where $W=\frac{3}{4}\openone-\ket{GHZ}\bra{GHZ}$, $\ket{GHZ}=(\ket{000}+\ket{111})/\sqrt{2}$  \cite{PhysRevLett.87.040401}. 
	From the measured death and revival evolution, we fit with the result of non-Markovianity as $\mathcal{M}=0.99999(1)$, which illustrates strong memory effect in Fig. \ref{fig-1}(a).

	%In Fig. \ref{fig-1}(b), the steering witness (SW) are considered in 1SDI and 2SDI scenarios \cite{gehring2015implementation}. 
	Due to the symmetric property of system state $\rho_{\mathrm{ABC}}(\theta=\pi/4,\eta=1)$, the corresponding reduced bipartite states are identical, namely, $\rho_{AB} = \rho_{AC} = \rho_{BC}$. In 1SDI scenario, the steerability from Alice to Bob and Charlie' s reduced state is quantified as $\mathcal{S}_{a\rightarrow bc}=1+0.1547\langle Z_BZ_C\rangle-(\langle A_2Z_B\rangle+\langle A_2Z_C\rangle+\langle A_0X_BX_C\rangle-\langle A_0Y_BY_C\rangle-\langle A_1X_BY_C\rangle-\langle A_1Y_BX_C\rangle)/3$, where $A_i$ for $i=1, 2, 3$ is the observables on Alice's side with outcomes labeled $\pm 1$, and Pauli operators $X$, $Y$ and $Z$ on Bob and Charlie's sides. In 2SDI scenario, the steerability from Bob and Charlie to Alice's reduced state is $\mathcal{S}_{bc\rightarrow a}=1-0.1831(\langle B_2C_2\rangle+\langle ZB_2\rangle+\langle ZC_2\rangle)+0.2582(-\langle XB_0C_0\rangle+\langle YB_0C_1\rangle+\langle YB_1C_0\rangle+\langle XB_1C_1\rangle)$, where $B_i$ and $C_i$ ($i=1, 2, 3$) are representing observables on Bob's and Charlie' s sides with $\pm 1$ outcomes and Pauli operators $X$, $Y$ and $Z$ \cite{gehring2015implementation}. 
	As shown in the main text Fig. \ref{fig-1}(b), we first fit the experimental results of the death process with theoretical analysis $f(x)=A*e^{-Bx^2}+C$ and obtain the fitting parameters $A=1.3911 (1.07412)$, $B=1.2999 (1.29996)*10^{-6}$ and $C=0.77026 (0.664544)$ for 1SDI (2SDI) scenario. Then applying these fitting results in the revival process, we further obtain the degree of non-Markovianity as $M=1 (1)$, which also verifies the strong memory effect. Moreover, the entangled yet unsteerable phenomenon, with the entanglement bound in the red dashed line, illustrates the hierarchy between entanglement and quantum steering. Our results indicate that the information previously lost to the environment flows back into the system due to the memory effect in non-Markovian process. The degree of non-Markovianity, which determines the recovery ability, can be operationally quantified through the whole death and revival dynamical processes of quantum correlations.

	\begin{figure}[t]
		\includegraphics[width=0.48\textwidth]{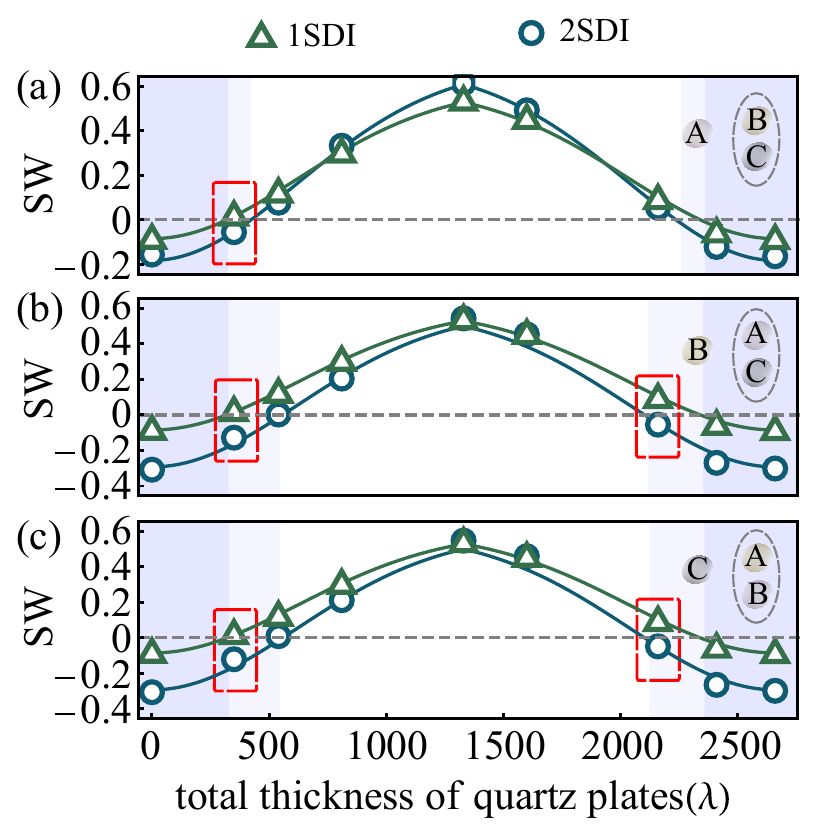}
		\caption{\textbf{Different configurations of non-Markovian evolution processes of system state ($\rho_{\mathrm{ABC}} (\theta=\pi/4,\eta=0.66)$).} (a) Alice versus Bob and Charlie ($\mathrm{A|BC}$). (b) Bob versus Alice and Charlie ($\mathrm{B|AC}$). (c) Charlie versus Alice and Bob ($\mathrm{C|AB}$). The horizontal axis is the total thickness of quartz plates $\mathcal{L}(\lambda)$, and the vertical axis is the degree of tripartite steering witness (SW), with the steerable boundaries in gray dashed lines. Hollow triangles and circles represent SW in one-sided and two-sided device independent scenarios (1SDI and 2SDI). The dark, light, and unshaded backgrounds are denoted states steerable in 1SDI and 2SDI scenarios, steerable only in 2SDI scenarios, and unsteerable in both regions. The states steerable only in 2SDI are highlighted in red dashed boxes. 
		}\label{fig-2}
	\end{figure}

	To reveal the hierarchical and directional dynamical features unique to multipartite systems, we further prepare the initial system state $\rho_{\mathrm{ABC}}(\theta=\pi/4,\eta=0.66)$ and investigate the whole non-Markovian evolution for three bipartitions respectively representing as  $(\mathrm{A|BC})$, $(\mathrm{B|AC})$ and $(\mathrm{C|AB})$ in both 1SDI and 2SDI scenarios, as shown in Fig. \ref{fig-2}. The noise term, controlled by the parameter $\eta$, is introduced to ensure the experimental distinguishability of different steering configurations in the 1SDI and 2SDI scenarios. Despite with the above SW calculations in bipartition ($\mathrm{A|BC}$), the tripartite quantum steerability in other bipartitions, ($\mathrm{B|AC}$) and ($\mathrm{C|AB}$), are also considered in two 1SDI and 2SDI scenarios, as illustrated in Fig. \ref{fig-1}. Our experimental results fit well with the theoretical analysis, where the dynamical evolution starts from steerable in both 1SDI and 2SDI scenarios to steerable only in the 2SDI scenario and unsteerable in both scenarios under the system-environment interaction $\mathrm{U}_1$. Subsequently, as the accumulation of the operation $\mathrm{U}_2$, the lost steerability is revival due to the memory effect and the system state returns to steerable in both 1SDI and 2SDI scenarios. Different from one-way steerable case in bipartite cases (for example, only Alice can steer reduced Bob's state), the states only steerable in one direction are more intricate, i.e., only Bob and Charlie can steer reduced Alice's state in (a), only Alice and Charlie can steer reduced Bob's state in (b) and only Alice and Bob can steer reduced Charlie's state in (c), as highlighted by the red dashed boxes. Notably, due to the asymmetric role of subsystem $A$ in system state $\rho_{\mathrm{ABC}}^s$, the critical effective thicknesses $\mathcal{L}(\lambda)$ at which the transitions from unsteerable to the steerable are different in Fig. \ref{fig-2}(a), whereas those in Figs. \ref{fig-2}(b) and (c) are the same. These observations show that multipartite quantum  steering exhibits richer dynamical behaviors than bipartite systems, including the hierarchical configurations and asymmetric revival patterns that are essential for designing robust quantum networks operating under practical environmental conditions. 	
	
	\emph{Conclusion.}---In this work, we experimentally observe the non-Markovian evolution of tripartite quantum steering, including the death and revival processes. Importantly, we investigate the hierarchical structure of tripartite quantum steering through different bipartitions to observe the asymmetric property of tripartite quantum steering. We also demonstrate that unsteerable yet entangled states provide direct evidence of the hierarchical relationships between entanglement and steering. Our results highlight the potential advantages of applying multipartite quantum system in the non-Markovian process with memory effect, and offer practical insights for leveraging steering resources in quantum networks. 
	
	Compared with the bipartite case, our results reveal the more intricate directional and hierarchical features of multipartite steering in non-Markovian environments and establish an experimentally accessible framework for studying multipartite quantum correlations in open systems, advancing the study of leveraging steering resources in quantum networks and 1SDI protocols under realistic noise conditions \cite{PhysRevLett.125.020404,deng2021sudden}. More broadly, our experimental platform enables precise control of the degree of system-environment interactions and provides a versatile setting for future investigations of quantum-resource dynamics in complex open systems \cite{RevModPhys.91.025001}.
	
	This work was supported by the National Natural Science Foundation of China Grants No. 12404403, Zhejiang Provincial Natural Science Foundation of China Grant No.LQN25A040018, Hangzhou Leading Youth Innovation and Entrepreneurship Team project Grant No. TD2024005, the National Key Research and Development Program of China (Grant No. 2024YFA1408900), the National Natural Science Foundation of China (Grant No. U21A20436), the Innovation Program for
	Quantum Science and Technology (Grant No. 2021ZD0301705)

	\bibliography{reference}% Produces the bibliography via BibTeX.
	\bibliographystyle{naturemag}

\end{document}